\documentclass[a4paper,12pt]{iopart}

\usepackage{epsfig}

\begin{document}

\title{Searching for hexagonal analogues of the half-metallic half-Heusler
      XYZ compounds}

\author{Frederick Casper and Claudia Felser} \address{Institut f\"ur Anorganische
Chemie und Analytische Chemie\\ Johannes Gutenberg-Universit\"at, Staudinger
Weg 9, D-55099 Mainz, Germany.}

\author{Ram Seshadri}  \address{Materials Department and Materials Research
Laboratory\\ University of California, Santa Barbara, CA 93106, USA.}

\author{C. Peter Sebastian and Rainer P\"ottgen} \address{Institut f\"ur
Anorganische und Analytische Chemie\\ Westf\"alische Wilhelms-Universit\"at
M\"unster\\ Correnstrasse 30, D-48149 M\"unster, Germany.}

\begin{abstract}

The \textit{XYZ} half-Heusler crystal structure can conveniently be 
described as a tetrahedral zinc blende \textit{YZ} structure which is 
stuffed by a slightly ionic \textit{X} species. This description is well
suited to understand the electronic structure of semiconducting 8-electron
compounds such as LiAlSi (formulated Li$^+$[AlSi]$^-$) or semiconducting
18-electron compounds such as TiCoSb (formulated Ti$^{4+}$[CoSb]$^{4-}$). The
basis for this is that [AlSi]$^-$ (with the same electron count as  Si$_2$)
and [CoSb]$^{4-}$ (the same electron count as GaSb), are both structurally
and electronically, zinc-blende semiconductors. The electronic structure of
half-metallic ferromagnets in this structure type can then be described as
semiconductors with stuffing magnetic ions which have a local moment: For
example, 22 electron MnNiSb can be written  Mn$^{3+}$[NiSb]$^{3-}$. The
tendency in the 18 electron compound for a semiconducting gap
-- believed to arise from strong covalency -- is  carried over in MnNiSb to a
tendency for a gap in one spin direction.  Here we similarly propose the
systematic examination of 18-electron hexagonal compounds for 
semiconducting gaps;  these would be the ``stuffed wurtzite'' analogues of the
``stuffed zinc blende'' half-Heusler compounds. These semiconductors could then serve
as the basis for possibly  new families of half-metallic compounds, attained
through appropriate  replacement of non-magnetic ions by magnetic ones. These
semiconductors and semimetals with tunable charge carrier concentrations could
also be interesting in the context of magnetoresistive and thermoelectric 
materials.  


\end{abstract}

\maketitle

\section{Introduction}

Half-metallic
ferromagnets \cite{deGroot:PRL.50.2024,Felser:Angewandte.46.668}, in contrast
to more usual ferromagnets, are completely spin-polarized, possessing a gap
in one spin direction at the Fermi energy $E_{\rm F}$. Recent interest in the
idea  that solid state devices can function through manipulation of the spin of
electrons \cite{Wolf:Science.294.1488,Zutic:RMP.76.323} has given rise to a
wealth of research in the area of half-metals, which play an important 
r\^ole in 
spin-injection \cite{Pickett:PhysToday.54.34,Grunberg:PhysToday.54.31}.
An important fundamental question is what makes some ferromagnets
half-metals, whilst others are not. This question has been addressed 
across large classes of materials such as the system 
Fe$_{1-x}$Co$_s$S$_2$ \cite{Mazin:APL.77.3000,Ramesha:PRB.60.214409}, in
proposed epitaxial transition metal compounds with the zinc blende 
structure \cite{Pask:PRB.67.224420}, in some chromium chalcogenide 
spinels \cite{Horikawa:JPhysC.15.2613,Park:PRB.59.10018}, and in the 
half-Heusler \cite{deGroot:PRL.50.2024,Galanakis:PRB.66.134428,Jung:Theochem.527.113,Kandpal:JPhysD.39.776,Kohler:InorgChem2007} and
Heusler \cite{Palmstrom:MRSBull.39.776,Galanakis:PRB.66.174429,Block:JSSC.176.646} compounds, 
CrO$_2$ \cite{Sorantin:InorgChem.31.567,Mazin:PhysRevB.59.411} and
some members of the colossal magnetoresistive 
manganites \cite{Singh:PhysRevB.53.1146}. With the exception of the perovskite
manganites, in all these different classes of half-metallic compounds,
an interesting common theme that emerges is the existence of a band semiconductor 
that is quite proximal in terms of composition and electron count. 

As an example, in Fe$_{1-x}$Co$_s$S$_2$ perhaps the  first series of compounds 
that were reported with integer moments on the magnetic 
substituents \cite{Jarrett:PhysRevLett.21.617}, the starting point is 
semiconducting FeS$_2$ whose empty $e_g$ band is populated through Co 
substitution.  Similarly, the basis for zinc blende
half-metals \cite{Pask:PRB.67.224420,Galanakis:PRB.67.104417} is the
replacement of cations in a semiconductor with magnetic ions:  Half-metallic
zinc blende CrAs can be considered the magnetic analogue of  GaAs, or perhaps
even more appropriately, as the magnetic analogue of zinc blende ScAs, with
the semiconducting gap being retained in the magnetic compound albeit in one
spin direction. Similar analogies can be drawn for systems such as rutile 
CrO$_2$ \cite{Sorantin:InorgChem.31.567,Mazin:PhysRevB.59.411}. 

One of the best studied systems of half-metals are the half-Heusler compounds
\textit{XYZ} exemplified by MnNiSb \cite{deGroot:PRL.50.2024}. Whangbo and
coworkers pointed \cite{Jung:Theochem.527.113} out that the  18 electron
half-Heusler compounds must be non-magnetic and  semiconducting. Recently,
Galanakis \cite{Galanakis:PRB.66.134428} has placed these half-Heusler
compounds on a firm theoretical footing, suggesting that in the half-metallic
compositions, the magnetic moment obtained from the saturation magnetization
$M$, per formula unit, should vary as $M = Z_t - 18$ where $Z_t$ is the total
number of  valence electrons. Some of us \cite{Kandpal:JPhysD.39.776} have 
examined the role that covalency plays and have proposed that the half-Heusler compounds
are both structurally and electronically best treated as an \textit{X} ion 
stuffing a zinc blende \textit{YZ} structure. When the \textit{X} ion is nonmagnetic
and  $Z_t$ =  18, the compound is a band semiconductor. If \textit{X} is 
magnetic and $Z_t \ne 18$, such as in MnNiSb with $Z_t = 22$, the compound 
is a half-metallic ferromagnet, with, in the case of MnNiSb, a magnetic moment 
of 4\,$\mu_B$ \textit{per} formula unit. 

In this contribution, we ask the following question: If \textit{XYZ} 
stuffed zinc blende compounds with $Z_t$ = 18 are semiconducting, can similar
18 electron semiconductors be found amongst 18 electron \textit{XYZ} stuffed
wurtzites, given that wurtzite and zinc blende are simply stacking variants
of one another with very similar topologies and hence bonding patterns. We
suggest  the answer is yes, and examine different closely related hexagonal \textit{XYZ} 
compounds: ScCuSn, LaCuSn and YCuSn. We use this to propose related
magnetic  compounds that would potentially comprise a novel class of  half
metallic \textit{XYZ} compounds with hexagonal crystal structure.

Hexagonal \textit{XYZ} compounds with cerium, europium, ytterbium, and 
uranium as the \textit{X} atom have been investigated in the last 
twenty years in light of their unusual properties. Examples include 
valence-fluctuations in EuPtP \cite{Lossau:ZPhysB.74.227}, the Verwey
type transition in EuNiP \cite{Ksenofontov:EPL.74.672}, intermediate-valent
YbCuAl \cite{Poettgen:HPCRE.32.453}, the  10\,K ferromagnet
CeAuGe \cite{Poettgen:JMMM.152.196}, the Kondo system 
CePtSn \cite{Riecke:ZNat.60b.821}, and  the heavy fermion material
CePtSi \cite{Lee:PRB.35.5369}. CeRhAs is a Kondo semiconductor in the stuffed
wurtzite structure, which undergoes an electronic transition at high
temperature and pressure into a metallic phase simultaneously with a structural
transition into the TiNiSi structure \cite{Umeo:PRB.71.064110}. The 
rare-earth--Pd--Sb
system is particularly interesting since CePdSb is a 17\,K Kondo ferromagnet 
with a resistance minimum, while many members of this series prepared with 
other magnetic rare-earths are antiferromagnetic \cite{Malik:JMMM.102.42}. 
The tunability of electronic structure and the charge carrier density in
this structure type make the compounds of this structure type interesting for
magnetoresitance effects \cite{Casper:ZAAC.632.1273,Pierre:JMMM.217.74},
and it is likely that they would be a fertile class of thermoelectric 
compounds as well.

\section{Computational methods}

Density functional theory-based electronic structure calculations reported
here were performed using the linear muffin tin orbital 
method \cite{StuttgartLMTO} within the local spin density approximation. The
crystal structure inputs for the calculations were obtained from 
experimental data in the literature, except when hypothetical structures
are considered. Crystal orbital Hamiltonian populations
(COHP) \cite{Dronskowski:JPhysChem.97.8617} and the electron  localization
function (ELF) \cite{Becke:JChemPhys.92.5397,Silvi:Nature.371.683}  were used
to obtain insights into bonding in the title compounds, respectively,  in
terms of the strengths of individual bonds, as well as in real space.

\section{Results and discussion}

\subsection{Crystal structures of the hexagonal compounds \textit{X}CuSn}

\begin{figure} 
\centering \epsfig{file=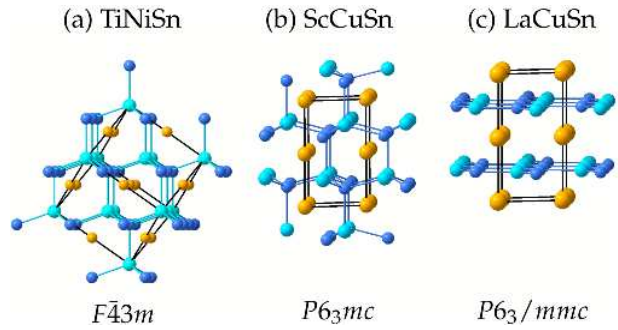, width=8cm}
\caption{(Color online) Crystal structures of (a) the half-Heusler compound
TiNiSn, (b) the stuffed wurtzite ScCuSn with the LiGaGe structure, and 
(c) LaCuSn, in the ZrBeSi structure type. The structures are depicted 
in a manner that highlights respectively, the zinc-blende NiSn, the
wurtzite CuSn, and the ``decorated graphite" CuSn networks. The orange spheres
are respectively Ti, Sc, and La. Light and dark blue spheres represent (Ni/Cu) 
and Sn.} 
\end{figure}

\begin{table}
\begin{tabular}{lllllll}
\hline
compound & Space Group & $a$ (\AA) & $c$ (\AA) & $c/a$ & $z$(Cu) & $z$(Sn) \\  
\hline 
LaCuSn	 & $P6_3/mmc$ & 4.583  & 8.173  & 1.783 & 0.75	  & 0.25\\ 
YCuSn    & $P6_3mc$   & 4.513  & 7.274  & 1.612 & 0.8148  & 0.2318\\ 
ScCuSn   & $P6_3mc$   & 4.388  & 6.830  & 1.557 & 0.82545 & 0.22914\\ 
\hline
\multicolumn{6}{c}{hypothetical structures}\\ 
\hline
LaCuSn [LiGaGe]   & $P6_3mc$ & 4.583  & 8.173  & 1.783 & 0.78    & 0.22\\ 
LaCuSn [wurtzite] & $P6_3mc$ & 5.005  & 8.173  & 1.633 & 0.8125  & 0.1875\\ 
\hline 
\end{tabular} 
\caption{Crystal structures of the compounds whose electronic structures
are described in this contribution. $X$ (Sc, Y, La)
at $(0,0,0)$, Cu at $(\frac 23, \frac 13, z\mbox{(Cu)})$ and 
Sn at $(\frac 13, \frac 23, z\mbox{(Sn)})$.
The first three experimental crystal
structures are taken from reference \cite{Sebastian:SSS}.} 
\end{table}

A huge variety of the equiatomic intermetallic compounds \textit{XYZ} 
(\textit{X} = rare earth, \textit{Y} = late transition metal element,
\textit{Z} = main group element) crystallize in structure types related to 
the AlB$_2$ family. The ordered superstructures crystallize in the LiGaGe, 
NdPtSb, and ZrBeSi type structures. The late transition metals and the main 
group elements form \textit{Y}$_3$\textit{Z}$_3$ hexagons,
which are connected in a two dimensional honeycomb network. Disorder between
the transition metals and the main group elements leads to the pseudobinary
structure types like AlB$_2$, Ni$_2$In, or CaIn$_2$. The
layers can be planar like in graphite (found in the ZrBeSi and AlB$_2$ types), 
weakly puckered (NdPtSb type) or strongly puckered with short interatomic 
distances between the layers leading to a wurtzite-related structure with
a three dimensional network (LiGaGe type). Compared to the compounds 
with the stuffed zinc blende  structure, namely the  C$_{1b}$ half-Heusler
compounds, the LiGaGe structure type which is the focus of this contribution
has a free lattice parameter, the $c/a$ ratio, which should be 1.633 for the
ideal hexagonal wurtzite structure. Beside the variable $c/a$ ratio, the 
free $z$ parameter of the 2$b$ positions (see table\,1) allows different
degrees of puckering of the hexagons leading to structures that can vary almost
continuously from three dimensional to quasi-two dimensional, with anticipated
changes in electronic properties. Due to this reduction in symmetry in 
comparison with the half-Heusler compounds, a large variety of different structure 
types are possible as described above. The different
superstructures are related via group-subgroup relations as recently
reviewed \cite{Hoffmann:ZKristallogr.216.127}. The bonding features in
such materials have been discussed in several overviews \cite{Pottgen:JAlloysCompounds.23.170,Nuspl:InorgChem.35.6922,Landrum:InorgChem.37.5754,Bojin:HelvChimActa.86.1653,Bojin:HelvChimActa.86.1683,Gaudin:ChemMater:17.2693}. 
Of the compounds discussed here, ScCuSn and YCuSn are examples
for the puckered LiGaGe-type structure, whereas LaCuSn is an example of 
a planar [YZ]$^{3-}$ (here CuSn$^{3-}$) network \cite{Sebastian:SSS}. 
CeCuSn is dimosphic with different degrees of puckering in the
low- and high-temperature modifications \cite{Sebastian:ZN}.

Figure\,1 compares the crystal structure of a typical cubic half-Heusler 
compound, (a) $F\bar{4}3m$ TiNiSn, with the crystal structures of two 
variants of the hexagonal \textit{XYZ} compounds discussed here: $P6_3mc$
ScCuSn \cite{Sebastian:SSS}, crystallizing in the LiGaGe type structure
and the planar $P6_3/mmc$ LaCuSn \cite{Sebastian:SSS}.  TiNiSn is
displayed with bonds connecting the zinc blende network of Ni and Sn. The Ti
atoms stuff the (6+4)-coordinate voids of the zinc blende structure. The
structure is displayed with the [111] direction pointing up in the plane of the
page, in order to emphasize the ABC stacking each of the three \textit{fcc}
substructures  of the half-Heusler structure. ScCuSn can be thought of as
comprising a Sc$^{3+}$ ion stuffing a wurtzite [CuSn]$^{3-}$ substructure
where the number of electrons is 18 (d$^0$ + d$^{10}$ + s$^2$ + p$^6$). 
LaCuSn is also a 18 valence electron compound which crystallizes in 
the ZrBeSi-type structure. In this structure type, compounds with usually 
18 valence electrons are found with some exceptions.

From the described electron count of these 18 electron compounds, we would
expect closed shell species similar to the 18-electron half-Heusler compounds: 
Non-magnetic, and semiconducting, at least for 	ScCuSn. In LaCuSn the
[CuSn]$^{3-}$ honeycomb network adopts a planar graphite type layer. This
structural change from ScCuSn \textit{via} YCuSn to LaCuSn is reflected in 
the $c/a$ ratio and the $z$ parameter, listed in table\,1. The $c/a$ ratio 
decreases with an increase of the puckering.    

\subsection{Electronic structures of the hexagonal compounds \textit{X}CuSn}

\begin{figure}
\centering \epsfig{file=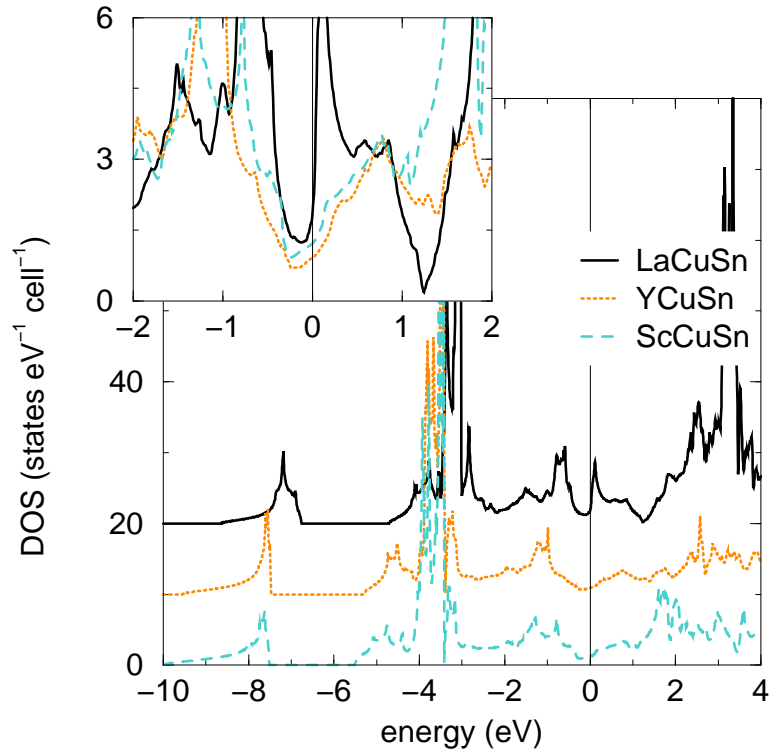, width=10cm}
\caption{(Color online) Total densities of state obtained from LMTO 
calculations on the experimental crystal structures of LaCuSn, YCuSn, and 
ScCuSn. The DOS are offset on the ordinate for clarity. The inset shows the 
DOS of the three compounds in a region close to $E_F$. These data are not
offset.}
\end{figure}

\begin{figure}
\centering \epsfig{file=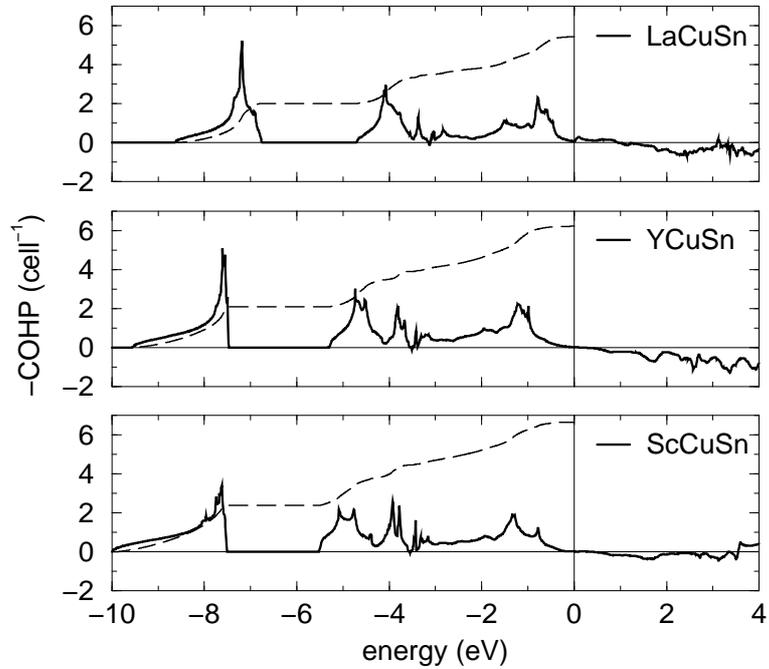, width=10cm}
\caption{Crystal orbital Hamiltonian populations (COHPs) of the Cu-Sn
interaction in the three compounds, LaCuSn, YCuSn, and ScCuSn.}
\end{figure}

We have examined the \textit{X}CuSn compounds using LMTO calculations. The
experimental lattice constants and experimental $z$-parameters were employed
in the input crystal structures to investigate the influence of puckering 
of the honeycomb \textit{Y}$_3$\textit{Z}$_3$ networks on the
electronic structure.  Additionally, calculations were performed for
LaCuSn in two hypothetical structures described in table\,1: The LiGaGe 
structure  with experimental lattice parameters and hypothetical $z$ 
parameters which result in puckered Cu$_3$Sn$_3$ honeycombs, and
an ideal wurtzite structure with the experimental $c$ parameter, and
$a$ and $z$ chosen so that the ideal wurtzite structure is obtained 
($c/a$ = 1.633). In this ideal structure, all CuSn$_4$ and SnCu$_4$ tetrahedra
are regular.

In the three panels of figure\,2, we compare the densities of states (DOS)
near the Fermi energy $E_F$ for LaCuSn, YCuSn, and ScCuSn calculated for
the experimental structures. We see trends in the electronic structure 
as the CuSn networks in these structures are increasingly puckered.   
All compounds have a noticeable pseudo band gap at $E_F$, with a more pronounced
gap in the more puckered Y and Sc compounds. Projections of the densities of 
state on the different orbitals (not displayed) reveal that the valence band
has mainly Sn p character.  The Sn s states are separated by a gap from the
valence band at around $-$8\,eV. The  bottom of the valence band is built from 
Sn p states. Cu d states lead  to spikes in the electronic structure around 
$-$3.5\,eV (ScCuSn) or $-$3\,eV (LaCuSn). The Cu d bands are very
flat with an over all dispersion of 0.5 eV. In the three-dimensional compound
ScCuSn all Sn p states contributes to the states  below and above the copper
states. In two dimensional LaCuSn, the bottom  of the valence states around 
$-$4\,eV has pronounced Sn p$_z$ character, whereas the top of the valence 
band is mainly built from Sn p$_x$ and p$_y$ states. Sc or rare-earth d states 
form the conduction band with a band width around 5\,eV and have a small 
contribution  to the Sn p states below $E_F$. From the density of states we 
can conclude that a description of the compound as \textit{X}$^{+3}$ ion 
stuffing a wurtzite or ``decorated graphite'' [CuSn]$^{3-}$ substructure is 
appropriate.  The total DOS of these hexagonal compounds look very similar to 
those of the half-Heusler compounds \cite{Kandpal:JPhysD.39.776}, with some 
differences in the projected densities due to the different electronegativities
of the constituent elements. 

The similarities and differences in the electronic structure of the three
compounds are further emphasized through an analysis of the Cu-Sn COHPs of 
the three stannides displayed in figure\,3. Cu-Sn is the main bonding
interaction in these compounds and bonding and antibonding states are
separated by the Fermi energy. The dashed line in this figure is an
integration of the COHP up to the $E_F$, yielding a number  that is
indicative of the strength of the bonding. The extents of the bonding and the
antibonding states of the Cu-Sn COHPs is slightly larger in the puckered
compounds compared to the planar LaCuSn. Integrating the COHPs, we find that
the strongest bonding interaction is in the planar LaCuSn. The planar Cu-Sn
layers seem to lead to a stronger Cu-Sn interaction than the puckered Cu-Sn
layers. However, the total bonding interaction seems to be stronger in the
puckered compounds. Cu-Sn interactions are nonbonding around and above $E_F$ 
and therefore do not contribute to the states around $E_F$. In YCuSn and 
ScCuSn, the other pairwise interactions are significant smaller: The 
\textit{X}-Cu interaction is only 10\%, and the \textit{X}-Sn interaction 
is around 20\% of the Cu-Sn interaction. The \textit{X}-Cu interaction is 
bonding below $E_F$ and small and slightly bonding above $E_F$. 
We can conclude that the bonding interactions in the hexagonal compounds 
ScCuSn and YCuSn are similar to what are found in the half Heusler 
compounds \cite{Kandpal:JPhysD.39.776}. A much stronger La-Cu interaction is 
found in the planar compound LaCuSn (not shown). The
La-Cu interaction leads to a strong bonding interaction leading to the high
density of states peak slightly above $E_F$. The La-Sn states 
build the top of the valence band and are 
responsible for reducing the magnitude of the pseudogap in this compound.
It is clear from this discussion that YCuSn and ScCuSn are electronically well
described as stuffed wurtzite compounds, with pseudogaps at $E_F$ between the 
bonding and antibonding states 

\subsection{The influence of puckering on the band structure and the
electron localization function}

\begin{figure}
\centering \epsfig{file=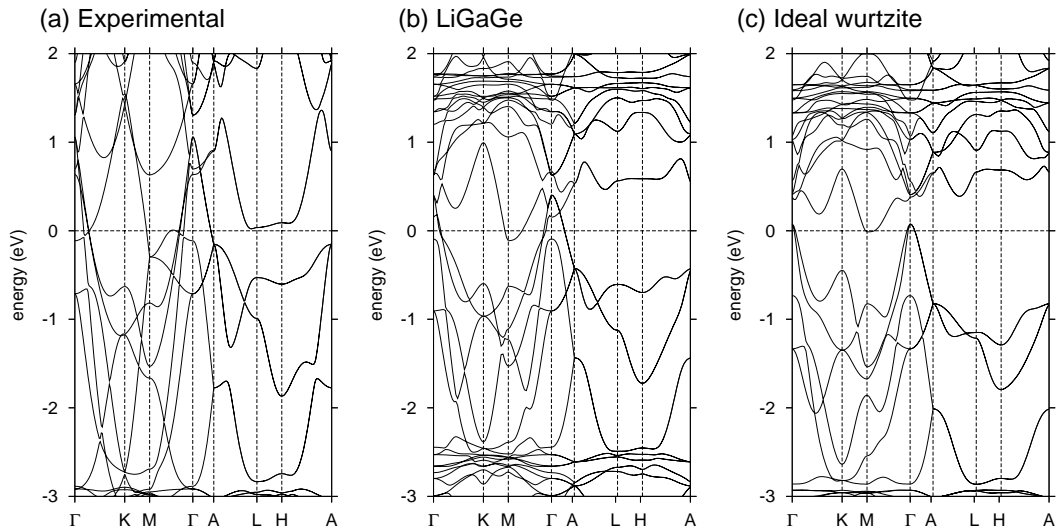, width=14cm}
\caption{Band structures of experimental and hypothetical LaCuSn 
structures.}
\end{figure}

The fact that the puckering of the Cu-Sn layer seems to be responsible for
the metallic or semimetallic behaviour of hexagonal 18 electron
compounds has motivated us to look more close on the influence of the
puckering on the band structure and the gap of these compounds.
Here we discuss the band structure of LaCuSn in the experimental structure 
and in the hypothetical LiGaGe-type and the wurtzite-type structures. 
Figure\,4 displays the band structure of (a) LaCuSn with the experimental 
ZrBeSi-type structure, (b) hypothetical LaCuSn with the LiGaGe-type and 
(c) LaCuSn with the ideal wurtzite-type structure. These structures are 
described in table\,1. LaCuSn in its experimental structure is clearly
a metal in the $\Gamma-K-M-\Gamma$ direction and shows a gap of nearly 
0.3\,eV along the direction at the surface of the Brillouin zone $A-L-H-A$. In the
LiGaGe structure the compound would be a semimetal due to the dipping of the
conduction band at the M point and the valence band at the $\Gamma$ point.
In the hypothetical wurtzite structure LaCuSn becomes semiconducting.
A closer view on the bands and their eigenvectors around $E_F$ shows that the
metallicity is due to an overlap between the conduction band with mainly
La d$_{x^2-y^2}$ and d$_{3z^2-1}$ character and the Sn p$_x$ and p$_y$ valence 
band.  Symmetry induces a metal to semiconductor transition by going from the
ZrBeSi-type structure to the LiGaGe-type structure. For the experimental
structure the La band dips below the Fermi energy at the $M$ point
with mainly d$_{3z^2-1}$ character and the Sn p$_x$ and p$_x$ bands cross 
$E_F$ around $\Gamma$. It is surprising that the overal band dispersion in the 
planar layers is much higher compared to the puckered layers. However a 
detailed look on the eigenvectors lead to a simple explanation. In case of 
the planar Cu-Sn layers the bands with Sn p$_x$ and p$_y$ contribution have a 
larger dispersion compared to the puckered layers. The strength of this 
interaction depends on the bonding distance and the effective overlap of the 
orbitals, both is stronger in case of the planar layers. The strong sigma type 
bonding interaction between Sn and Cu pushes the valence band above the Fermi 
level at the $\Gamma$ point. The conduction band which is derived from 
La d$_{3z^2-1}$ displays a dispersion of nearly 2\,eV and dips 
below $E_F$ along the $\Gamma$ to $M$ direction. 

The same band of the puckered structures, the La d$_{3z^2-1}$ band, has a
much smaller dispersion of 0.5\,eV and touches the Fermi energy slightly at
the $M$ point. The most important point is that the degeneracy at the $M$
point between the La d$_{3z^2-1}$ and the Sn p$_x$ band is lifted by reducing 
the symmetry from the $P6_3/mmc$ to $P6_3mc$, which renders the opening of the 
gap possible in the LiGaGe-type and wurtzite-type structure. In a manner 
similar to the half-Heusler compounds, the hexagonal compounds display an 
indirect gap, here between the $\Gamma$ and the M point.

By puckering the structure in going from the graphite type structure via
LiGaGe to the ideal wurtzite-type structure the overlap between the conduction
and the valence band becomes smaller and the symmetry is reduced. The
compounds with LiGaGe structure can show semiconducting or semimetallic
behaviour, what is experimentally and theoretically found for 
CePdSb \cite{Slebarski:JAlloysCompds.423.15} and 
CeRhAs \cite{Umeo:PRB.71.064110}.  Introducing $f$-elements like Gd lead to 
half metallic ferromagnets for example in GdPdSb \cite{Casper:JPhysD.40.3024}.

\begin{figure}
\centering \epsfig{file=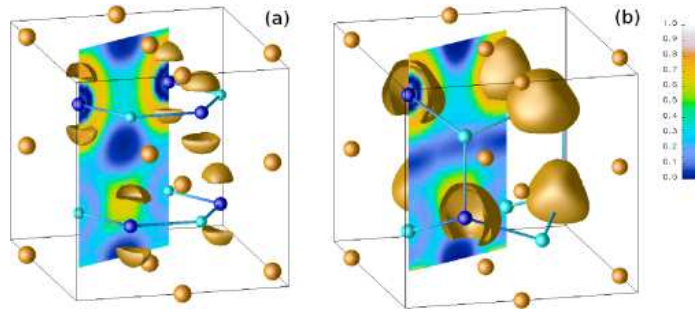, width=9cm}
\caption{(Color online) Electron localization functions (ELFs) of the valence 
electrons of LaCuSn in (a) the experimental crystal structure and in the (b) 
hypothetical wurtzite structure. The ELF isosurface is displayed for a value 
of 0.75.}
\end{figure}

\begin{figure}
\centering \epsfig{file=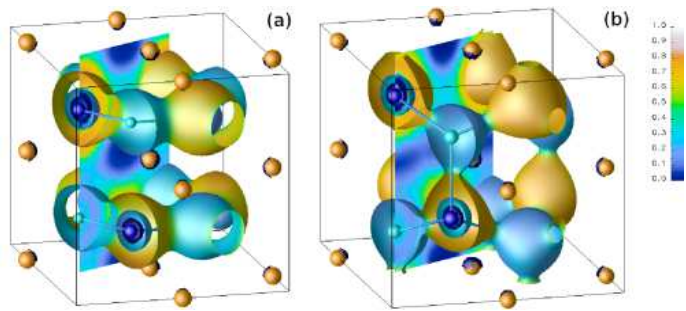, width=9cm}
\caption{(Color online) Electron density isosurfaces plotted for values
of 0.025\,$e$\AA$^{-3}$ for LaCuSn in (a) the experimental crystal structure 
and in the (b) hypothetical wurtzite structure. The electron density is 
colored according to the degree of localization.}
\end{figure}

Real space visualization of the electronic structure employed the electron
localization function (ELF) for LaCuSn in figure\,5(a) for the experimental 
ZrBeSi-type and in figure\,5(b) for the hypothetical wurtzite structure. 
The  value of localization runs from 0 (no localization, deep blue) to 1
(high localization, white). The isosurface of 0.75 is displayed here. The 
colored map in the background is the ELF shown on a (11$\bar2$0) plane to 
visualize the bonding interaction along the Cu-Sn network.  For
the ZrBeSi type structure figure\,5(b) the Sn p electrons perpendicular to 
planar Cu-Sn layer in are strongly localized. These p states interact strongly
with La d$_{3z^2-1}$ states with the layer, and are responsible for closure of 
the gap and the metallicity of the compound. In figure\,5(b), we find    
clear evidence for highly covalent and three-dimensional bonding between 
between Cu (cyan) and Sn (blue) in the wurtzite substructure of the wurtzite-type
LaCuSn. As observed in the half Heusler compounds, the localization is closer 
to the more electronegative Sn atom \cite{Kandpal:JPhysD.39.776}. 

The valence charge densities displayed in figures\,6(a) and (b)  are again very
similar to the valence charge densities of the half Heusler 
compounds \cite{Kandpal:JPhysD.39.776}. Because of the filled d shells on Cu 
it forms large nearly spherical blobs around that atom, visualized for a 
charge density of 0.025\,$e$\AA$^{-3}$. We find that these blobs of charge 
are pulled out into four strongly localized (as seen from the colouring) lobes 
arranged tetrahedrally and facing Sn in (b), whilst in the LiGaGe structure,
the bonding interaction in the third dimension is removed, leading to three
lobes arranged trigonal in the plane and facing the Sn in the plane.      

\section{Conclusion}

We have carried out a systematic examination of \textit{XYZ} compounds with 
hexagonal structures: LaCuSn, YCuSn, and ScCuSn, and demonstrated that 
18-electron compounds can become semiconductors depending
on the degree of puckering of the [CuSn]$^{3-}$ substructure. We demonstrate 
that the most instructive way of considering these systems is to think of 
them as being built up of a wurtzite [\textit{YZ}] framework that is stuffed 
with the electropositive \textit{X}. An effective strategy for new half-metals, 
in analogy with half-Heusler half-metals, would be to stuff these wurtzite-derived 
structures with magnetic ions.

\ack{CPS, HCK, CF, and RP acknowledge support from the  DFG within the priority
programme SPP 1166 \textit{Lanthanoidspezifische Funktionalit\"aten in 
Molek\"ul und Material}. CPS also acknowledges the NRW Graduate School of 
Chemistry for a PhD stipend. RS acknowledges the US NSF for support through 
a Career Award (DMR04-49354).}

\bibliography{hexHMF} \bibliographystyle{iopart-num}

\end{document}